\documentstyle[prl,aps,epsfig]{revtex}

\begin{document}

%\draft
\twocolumn

\title{Internal Spatiotemporal Stochastic Resonance 
       in a Microscopic Surface Reaction Model}
\author{O. Kortl\"uke$^\ddagger$
        \footnote{Corresponding author; Electronic mail:
          O.Kortlueke@tu-bs.de,
          Tel: +49-531-391-5397,
          Fax: +49-531-391-4577},
        V.N.~Kuzovkov$^{\dagger\ddagger}$,
        and W. von Niessen$^\ddagger$}
\address{$^\ddagger$Institut f\"ur Physikalische und Theoretische Chemie,
         Technische Universit\"at Braunschweig, Hans-Sommer-Stra{\ss}e
         10, 38106 Braunschweig, Germany\\
         $^\dagger$Institute of Solid State Physics,
         University of Latvia, 8 Kengaraga Street, LV -- 1063 RIGA, Latvia\\
}
\date{Received \today}
\maketitle

\begin{abstract}
We show the existence of internal stochastic resonance in a 
microscopic stochastic model for the oscillating CO oxidation
on single crystal surfaces. This stochastic resonance arises
directly from the elementary reaction steps of the system without
any external input.
The lattice gas model is investigated by means of Monte Carlo
simulations. It shows oscillation phenomena and mesoscopic pattern
formation. Stochastic resonance arises once homogeneous nucleation
in the individual surface phases (reconstructed and non-reconstructed)
is added. This nucleation is modelled as a noise process.
As a result, synchronization of the kinetic oscillations is
obtained. Internal stochastic resonance may thus be an internal
regulation mechanism of extreme adaptability.
\end{abstract}

The term stochastic resonance (SR) is given to the somewhat
counter-intuitive phenomenon
that in a non-linear system a weak signal can be amplified by
the assistance of noise. It has been introduced 
in 1981 by Benzi et al.\cite{benzi8101,benzi8201} 
in the context of a study
about the periodically recurrent ice ages. Over the last two decades
it has continuously attracted increasing attention and was shown
to occur in many systems%
\cite{wiesenfeld9501,gammaitoni9801,luchinsky9801}
in biology\cite{douglass9301,moss9501,levin9601},
chemistry\cite{guderian9601,foerster9601,hohmann9601},
and physics\cite{fauve8301,mcnamara8801,buchleitner9801}.
Generally systems showing SR are described in a formal
mathematical way using phenomenological macroscopic equations
of the mean field type including (i) a bistable system with
an activation barrier or some sort of threshold, (ii) a weak
coherent input, and (iii) a strong external noise which helps 
to overcome the activation barrier.
These macroscopic equations in a sense are able to describe
many different systems (because stochastic resonance is a general
phenomenon which occurs in many natural systems)
and have been used in the description of stochastic resonance
phenomena in 
Nd-YAG lasers\cite{celet9801}
homogeneous\cite{dykman9501} 
as well as heterogeneous\cite{yang9801,yang9802}
chemical reactions,
bistable quantum systems\cite{thorwart9701}
or the Lotka-Volterra model\cite{vilar9802}.
More complex systems (e.g.~a 
summing network of excitable units\cite{collins9501},
sheep populations\cite{grenfell9801}, 
two-dimensional excitable media showing spatio-temporal
     pattern formation\cite{jung9501},
sensory systems in crayfish\cite{douglass9301} 
or in the visual cortex\cite{stemmler9501},
or neuron-like systems\cite{fakir9801})
are generally modeled via Langevin equations or the 
Fitzhugh-Nagumo model.

In addition a few special systems have been investigated
in a more general manner via a
macroscopic mathematical description, e.g.~an
autonomous oscillating system\cite{hu9301}, 
a system in the limit of weak noise\cite{shneidman9401},
a system showing stochastic multiresonance\cite{vilar9702},
and non-dynamical systems with both internal and external
noise\cite{vilar9801}. 
Note that in the latter case the
internal noise is modeled in the same way as an external
noise and that it is only regarded as a general
internal noise without specifying the physical background.
Computer simulations performed to date consider coupled
neurons or general threshold devices, which are mesoscopic
models, i.e.~the microscopic physical picture is again neglected.

Our present model system is very unusual in the research on SR.
It gives for the first time a microscopic description of the
phenomenon of internal stochastic resonance and demonstrates its
physical reasons on the microscopic (atomic) length scale.
It is based on stochastic transitions, each with a clear
physical meaning. Without noise, the system exhibits a spatially
extended heterogeneous stable state with inherent local oscillations.
More important, the noise is not an external input but corresponds
to a physically realistic internal nucleation process. 
Because of the clear physical picture on the microscopic (atomic)
level our model is of course specialized and cannot describe
a large variety of different systems.
But on the other hand the results of this model and the
conclusions which can be drawn are very general ones and
suggest that internal SR may be the reason for
inherent synchronization and cooperative phenomena
in many physical, chemical and biological systems.

We consider a slightly modified version of 
a previously presented model for the
catalytic CO+1/2 O$_2$ reaction on Pt single crystal
surfaces\cite{kuzovkov9801,kortlueke9901},
which shows different types of kinetic oscillations in agreement
with experimental results. 
The model involves CO adsorption, desorption and diffusion,
dissociative O$_2$ adsorption and two surface phases
(reconstructed and non-reconstructed) which form and
propagate governed by the coverage with CO. The details
are given below. An extended version of the model
for the CO+NO reaction on Pt(100) is able to describe the
experimentally observed transition into chaotical behavior via
the Feigenbaum route\cite{kortlueke9804}.
Our model follows the well known model by Ziff, Gulari and 
Barshad\cite{zgb8601}
(ZGB model) and is investigated by means of Monte Carlo (MC)
simulations.
The Pt(110) surface of the catalyst is represented by a square
lattice of side length $L$ and lattice constant $a=1$.
From experiment\cite{imbihl9501}
it is well known that kinetic oscillations are closely connected
with the $\alpha\rightleftharpoons\beta$ reconstruction of the
Pt(110) surface, where $\alpha$ and $\beta$ denote the
$1 \times 2$ and the $1 \times 1$ surface phase, respectively.
In our model CO is able to adsorb onto a free surface site 
with rate $y_{\rm CO}=y$ and 
to desorb from the surface with rate $k$, 
independent of the surface phase the site belongs to.
O$_2$ adsorbs dissociatively onto two nearest neighbor (NN)
sites with different sticking coefficients $s$ onto the
two phases ($s_\alpha=0.5$, $s_\beta=1$). Therefore we get
the oxygen adsorption rates $y^\alpha_{\rm O}=1-y$ 
and $y^\beta_{\rm O}=2(1-y)$ for the $\alpha$ and $\beta$ phase,
respectively. For O$_2$ adsorption directly at the phase border
where one site belongs to the $\alpha$ and the other one to the
$\beta$ phase the geometric mean of these adsorption rates is used.
In addition, CO is able to diffuse with rate $D$ via hopping
onto a vacant NN site.
The CO+O reaction occurs, if CO hops to a site which is covered by O
and the reaction product CO$_2$ desorbs immediately from the surface.
All these processes are associated with the above kinetic transition
rates of the stochastic model which therefore determine the relative
speed of the individual reaction processes.
In the present study we use $y=0.51$, $D=100$, and $k=0.1$ as standard
values because CO diffusion is by far the fastest process.
For details see 
refs.~\cite{kuzovkov9801,kortlueke9901,kortlueke9804,kortlueke9803}.

The $\alpha \rightleftharpoons \beta$ phase transition is modeled
as a linear phase border propagation. Consider two NN surface sites
in the state $\alpha\beta$. The transition
$\alpha\beta\rightarrow\alpha\alpha$
($\alpha\beta\rightarrow\beta\beta$)
occurs if none (at least one) of these two sites is occupied by CO.
This phase border propagation mechanism mimicks the growth of the
$\beta$ phase because of the larger adsorption energy of CO on the
$1 \times 1$ phase than on the $1\times 2$ phase\cite{imbihl9501}.
The individual phases are stable or metastable. The direct
transition from a globally homogeneous $\alpha$ phase into a
homogeneous $\beta$ phase (or vice versa) is impossible; the activation
barrier is infinite. The stability of the individual phases
depends on the chemical coverage $\Theta_i$ of species $i$
on the surface of the catalyst.
For $\Theta_{\rm CO}<0.3$ the $\alpha$ phase, for larger values the
$\beta$ phase is stable\cite{kortlueke9901}.
The coverages of CO and O vary in the course
of the reaction because of the different sticking coefficients of O$_2$
on the two surface phases. Starting with a heterogeneous
distribution of the $\alpha$ and $\beta$ phase the activation barrier
for the surface phase transition is finite, but the transition into a
globally homogeneous phase does not occur because of the finite
surface phase propagation velocity, i.e.~the $\alpha$ or $\beta$
phases cannot grow to macroscopic islands. Therefore the
oscillations remain local, interfere and cancel each
other on sufficiently large surfaces\cite{kuzovkov9801,kortlueke9901}.
The system exists in a heterogeneous, dynamically stable state with
oscillations, which are locally synchronized by CO diffusion but
disappear on the macroscopic length scale for large lattices.
The $\alpha$ and $\beta$ phase, however, build almost homogeneous
islands on a mesoscopic length scale.

The nucleation is modeled as a spontaneous $\alpha\rightarrow\beta$ or
$\beta\rightarrow\alpha$ transition of a single site, completely
independent of its neighbors or the particle adsorbed onto this site.
Therefore this nucleation is a homogeneous process which
corresponds to a weak noise which generates dynamic defects on the
surface. The term weak noise is
used because the nucleation rate $\gamma\in[10^{-6},10^{-2}]$
is for the relevant values
several orders of magnitude smaller than the other transition rates
which are of the order of $10^{-1}$ to $10^3$. The defects grow or vanish
via the $\alpha\rightleftharpoons\beta$ reconstruction depending
on their chemical neighborhood, i.e.~the presence or absence of CO.
It has to be emphasized that all phenomena such as local oscillations,
growth and decline of the heterogeneously distributed surface phase islands,
quasiperiodical and chaotical behavior exist even without the
consideration of the nucleation 
process\cite{kuzovkov9801,kortlueke9901,kortlueke9804}.
Because of the very small nucleation rate, which introduces only
very few defects into the existing heterogeneous surface state,
one might suppose that the defects could
have no influence on the oscillating system, but
we shall demonstrate that the nucleation process
may have a profound influence on the system behavior leading
to cooperative phenomena and the synchronization of the local
oscillations.

In comparison with the standard problem of SR there exist
several major differences.
(i)
We have a two-phase system with an additional chemical coverage
instead of a bi-stable system. In addition none of the
two phases becomes homogeneous. Both coexist in a dynamically
stable heterogeneous state.
(ii)
No external coherent input is considered in our model. The
local oscillations originate from the
kinetic definition of the model itself and are very small on
a macroscopic, global length scale. These internal oscillations
can be seen as a substitute of the commonly used coherent input.
The macroscopic synchronization of the local oscillations
then corresponds to a large output signal.
(iii)
The source of noise is an internal physical process of the system,
the homogeneous nucleation, which breaks the mesoscopic homogeneity
of the surface phase islands.
The noise process is therefore not an addition to a periodic
input but is independent of the oscillations.
In most previous studies a weak coherent input is coupled
with a strong external noise. 
(iv)
The model is investigated via MC simulations which
correspond most closely to a hierarchy of master equations
with all correlations included and then
mapped onto a finite lattice. The simulation procedure
contains additional noise by its very nature but such a noise does not
lead to SR.
We may characterize the structure of our model by saying that practically
all processes which give rise to SR (except for the particle flux to the
surface) are internal to the system.

\begin{figure}
  \begin{center}
    \epsfig{file=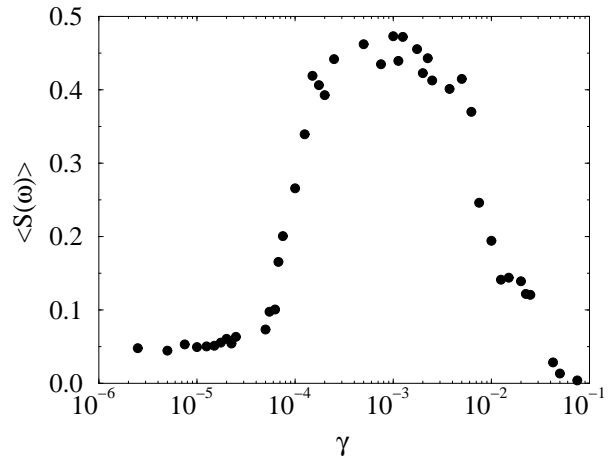,angle=270,width=8cm}
    \caption{
      Global amplitude $S(\omega)$ of the $\beta$ phase
      coverage as a function
      of the nucleation rate $\gamma$ which gives the noise strength. 
      A stochastic resonance maximum can be seen.
      The scatter of the points around the maximum are a result
      of the finite frequency interval due to FFT analysis.
      The values of the amplitude are averaged over 20 simulation
      results. All other parameters are kept constant at
      $y=0.51$ (CO adsorption), $k=0.1$ (CO desorption),
      $D=100$ (CO diffusion), $V=1$ (surface phase propagation),
      and $L=256$.
      }
    \label{fig:amp}
  \end{center}
\end{figure}

As can be seen in fig.~\ref{fig:amp} the nucleation (noise)
has a strong influence on the system behavior.
Without or with small nucleation rates $\gamma<5\cdot 10^{-5}$
only local oscillations exist. The amplitude of the
global oscillations vanishes for simulations on large
lattices\cite{kuzovkov9801,kortlueke9901}. 
With increasing nucleation rate
(strength of the noise) the local oscillations are synchronized on
a macroscopic scale and almost reach the theoretical maximum
of $S(\omega)=0.5$ ($\Theta_\beta$ varies between 0 and 1).
This holds also for larger lattices up to $L=4096$ which
is the current limit for our simulations.
Further increase of the noise decreases this synchronization until
at a nucleation rate of $\gamma=10^{-1}$ the system is completely
governed by strong noise. This behavior is the fingerprint of
SR\cite{gammaitoni9801}.

\begin{figure}
  \begin{center}
    \epsfig{file=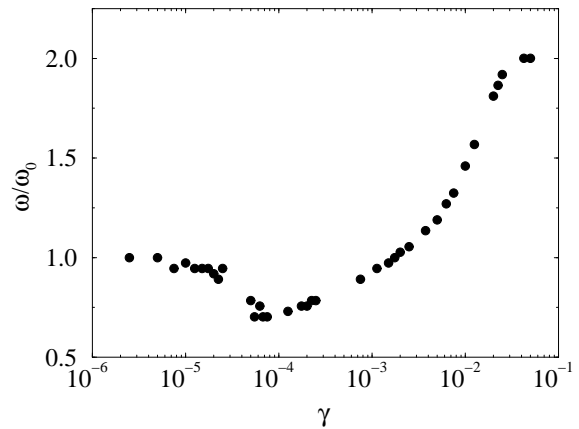,angle=270,width=7.6cm}
    \caption{
      Ratio of the frequencies $\omega/\omega_0$ as a function of 
      the nucleation rate $\gamma$ which gives the noise strength.
      Each value is averaged over 20 independent MC simulations.
      The errors are smaller than the symbol size. All other 
      parameters are kept constant at
      $y=0.51$ (CO adsorption), $k=0.1$ (CO desorption),
      $D=100$ (CO diffusion), $V=1$ (surface phase propagation),
      and $L=256$.
      }
    \label{fig:freq}
  \end{center}
\end{figure}

In addition the noise has an influence on the frequency of the
oscillations (see fig.~\ref{fig:freq}). The normal frequency of
the system is $\omega_0$, which can
only be observed in simulations on small lattices. For
nucleation rates $\gamma > 5\cdot 10^{-5}$ the phenomenon of
SR occurs. The nucleation forces the system to
oscillate with a different frequency $\omega$ starting at
$\omega \approx 2/3\, \omega_0$ for $\gamma=5\cdot 10^{-5}$. With
increasing nucleation rate the frequency increases
as well up to a value of about $\omega \approx 2\, \omega_0$ at
$\gamma=5\cdot 10^{-2}$.
This increase in the frequency is based on the increasing number of
dynamic phase defects which grow very fast and accelerate the 
corresponding phase transition. If the number
or density of defects becomes too large the oscillating behavior
breaks down.
The shape of the curve in fig.~\ref{fig:freq} in the interval
$\gamma\in [10^{-5},10^{-1}]$ is in very good agreement with the
one obtained in the study of a general autonomous system by
Haken et al.\cite{hu9301}.

\begin{figure}
  \begin{center}
    \epsfig{file=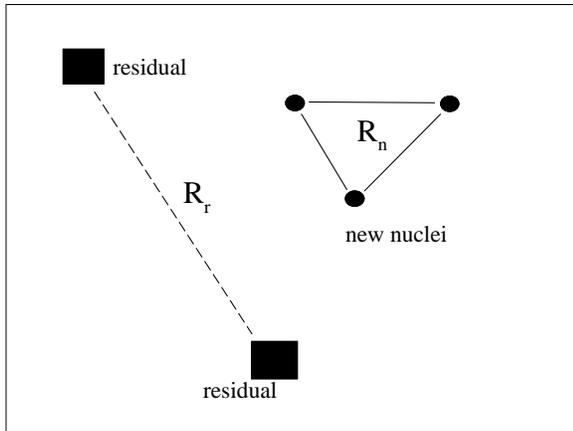,angle=270,width=7.6cm}
    \caption{
      New phase nuclei lead to a decreased mean distance $R_{\rm n}$
      between the individual phase islands. This then leads to
      global synchronization over the whole lattice. See text for details.
      }
    \label{fig:expl}
  \end{center}
\end{figure}

The underlying mechanism of this internal stochastic resonance
effect is unexpectedly simple. During the growth and decline
of the individual surface phases only a few very small residual
phase islands remain in domains where the local synchronization
leads to large amplitudes in the phase oscillations.
These residual islands are
spatially separated at a mean distance $R_{\rm r}$. Adsorbate
diffusion is well known to synchronize individual surface
domains within the socalled synchronization 
lentgh $\xi \propto \sqrt{DT}$\cite{kuzovkov9801,kortlueke9901},
where $D$ is the diffusion rate and $T$ is the time for one
oscillation period. Without nucleation $\xi < R_{\rm r}$
holds and only locally synchronized oscillations exist.
In addition, the residual islands can only combine into
a homogeneous phase if the second condition $R_{\rm r} \sim VT$
for the phase border propagation is fulfilled. But this condition
is also violated because $R_{\rm r} > VT$ holds.
Nucleation and subsequent growth now leads to new small phase 
islands (see fig.~\ref{fig:expl}).
The mean distance between the individual phase islands decreases
to $R_{\rm n}$, for which $R_{\rm n} < \xi < R_{\rm r}$ and
$R_{\rm n} \sim VT$ holds for proper nucleation rates. 
Therefore the separated phase islands can now be 
connected via island growth and
synchronized via CO diffusion. This results in
macroscopic synchronized oscillations.

The existence of the surface phase nucleation is well known
but has not been investigated experimentally yet,
in contrast to the initial growth of small surface phase
domains\cite{hopkinson9302}.
The nucleation rate should depend on the temperature
$\gamma=\gamma(T)$, but it is almost impossible to achieve an isolated
variation of the nucleation rate under experimental conditions
because all other parameter such as CO desorption and CO diffusion
also strongly depend on the temperature. 
It might thus be very difficult to experimentally
verify the mechanism behind the SR phenomenon in our model
but there should be other systems where it is feasible.

The nucleation of dynamic surface defects
as an internal process generates 
globally synchronized oscillations in our
CO+O$_2$/Pt(110) model reaction system via SR. Moreover,
because noise is always present in real systems,
this type of internal SR should be a very general phenomenon and
it may be the reason for cooperative phenomena and internal
synchronization via noise processes in many 
physical, chemical, and biological systems
where is has not been investigated experimentally
or even suspected to be present to date. This especially holds for
systems which exhibit inherent oscillations which are synchronized
on macroscopic length scales. In this case often noise effects
are supposed to be negligible, but as shown above they can also
be the origin of those cooperative phenomena. 
We believe that SR may thus be an internal regulation mechanism of
extreme adaptability. This conclusion
is drawn to search for internal SR in a variety of
experiments, because to date experiments designed to study the
role of internal noise have been inconclusive\cite{pantazelou9501}.

Financial support from the Deutsche Forschungsgemeinschaft,
the Volkswagen--Stiftung and the Fonds der chemischen
Industrie is gratefully acknowledged.

\end{document}